\documentclass{aastex}          
\usepackage{spr-astr-addons}    

\begin{document}
%
\title{MCP detector development for UV space missions}

\shorttitle{UV MCP Detectors}
\shortauthors{Conti et al.>}

\author{Lauro Conti} \and \author{J\"urgen Barnstedt} \and \author{Lars Hanke} \and \author{Christoph Kalkuhl} \and \author{Norbert Kappelmann} \and \author{Thomas Rauch} \and \author{Beate Stelzer} \and \author{Klaus Werner}
\affil{Institut f\"ur Astronomie und Astrophysik (IAAT), Universit\"at T\"ubingen, Sand 1, T\"ubingen, Germany}
\email{conti@astro.uni-tuebingen.de} 
\and 
\author{Hans-Rudolf Elsener}
\affil{Empa, Swiss Federal Laboratories for Materials Science and Technology, Ueberlandstrasse 129, D\"ubendorf, Switzerland}
\and
\author{Daniel M. Schaadt}
\affil{Institute of Energy Research and Physical Technologies, Clausthal University of Technology, Leibnizstra\ss{}e 4, Clausthal-Zellerfeld, Germany}


\begin{abstract}
We are developing imaging and photon counting UV-MCP detectors, which are sensitive in the wavelength range from far ultraviolet to near ultraviolet. A good quantum efficiency, solar blindness and high spatial resolution is the aim of our development. 
The sealed detector has a Cs-activated photoactive layer of GaN (or similarly advanced photocathode), which is operated in semitransparent mode on (001)-MgF$_2$. The detector comprises a stack of two long-life MCPs and a coplanar cross strip anode with advanced readout electronics. The main challenge is the flawless growth of the GaN photocathode layer as well as the requirements for the sealing of the detector, to prevent a degradation of the photocathode. 
We present here the detector concept and the experimental setup, examine in detail the status in the production and describe the current status of the readout electronics development.
\end{abstract}


%

\section{Introduction}\label{s:Introduction}

In the "Institut f\"ur Astronomie und Astrophysik T\"ubingen" (IAAT) we have long-term experience in developing and building imaging and photon counting UV microchannel plate (MCP) detectors, with the Echelle detector for two ORFEUS flights as our most remarkable success in this field \citep{refId0}.
The designs presented here are our current detector with Cs$_2$Te photocathode and the next generation which is developed for a 50~cm telescope for the European Stratospheric Balloon Observatory Design Study (ESBO DS) which will start in early 2018.
This mission will be a proof-of-concept and set the basis for further space missions with our detector. Due to the advancements in technology in nearly every part and especially for the readout electronics, our current and next generation detector is a small and lightweight MCP detector with very low power consumption. That means that its mass will be below 3~kg for the detector, its electronics and the high voltage supply at a power consumption of 10-15 Watt \citep{doi:10.1117/12.2054742}.
We chose to build an MCP detector as they are solar blind by design (in contrast to a CCD), they provide fast photon counting and a Poissonian noise dominated only by photon statistics \citep{doi:10.1063/1.1136710}. They are radiation hard and do not need cooling \citep{7814337}.

Right now, the only space observatory available which can obtain high resolution spectra is the Hubble Space Telescope (HST). It could reach its mission end in the near future with WSO-UV as the only successor in the next years \citep{2016ARep...60....1B}.
Here our detector could come into play: Smaller missions with short development period could close the gap until new flagship missions (e.g. LUVOIR) have their first light. Our detector would easily fit into a micro-satellite mission and be ready for a flight within two to three years. 

With cutbacks in spectral resolution and photon flux even a CubeSat mission would be possible with a size of at least 6U $\hat{=}$ 10x20x30~cm as described in \cite{Brosch2014}.

For the ESBO DS mission our goal is to use GaN as photocathode material in semitransparent mode on MgF$_2$. Dependent on the desired wavelength range for a specific mission, other photocathode and window materials could be used. With a door opened in space, instead of a sealing window, even spectra in the far ultraviolet spectral range (FUV, 92-120~nm) could be obtained.~


\section{Detector Design}\label{s:Design}

\begin{figure}[t]
\includegraphics[width=\columnwidth]{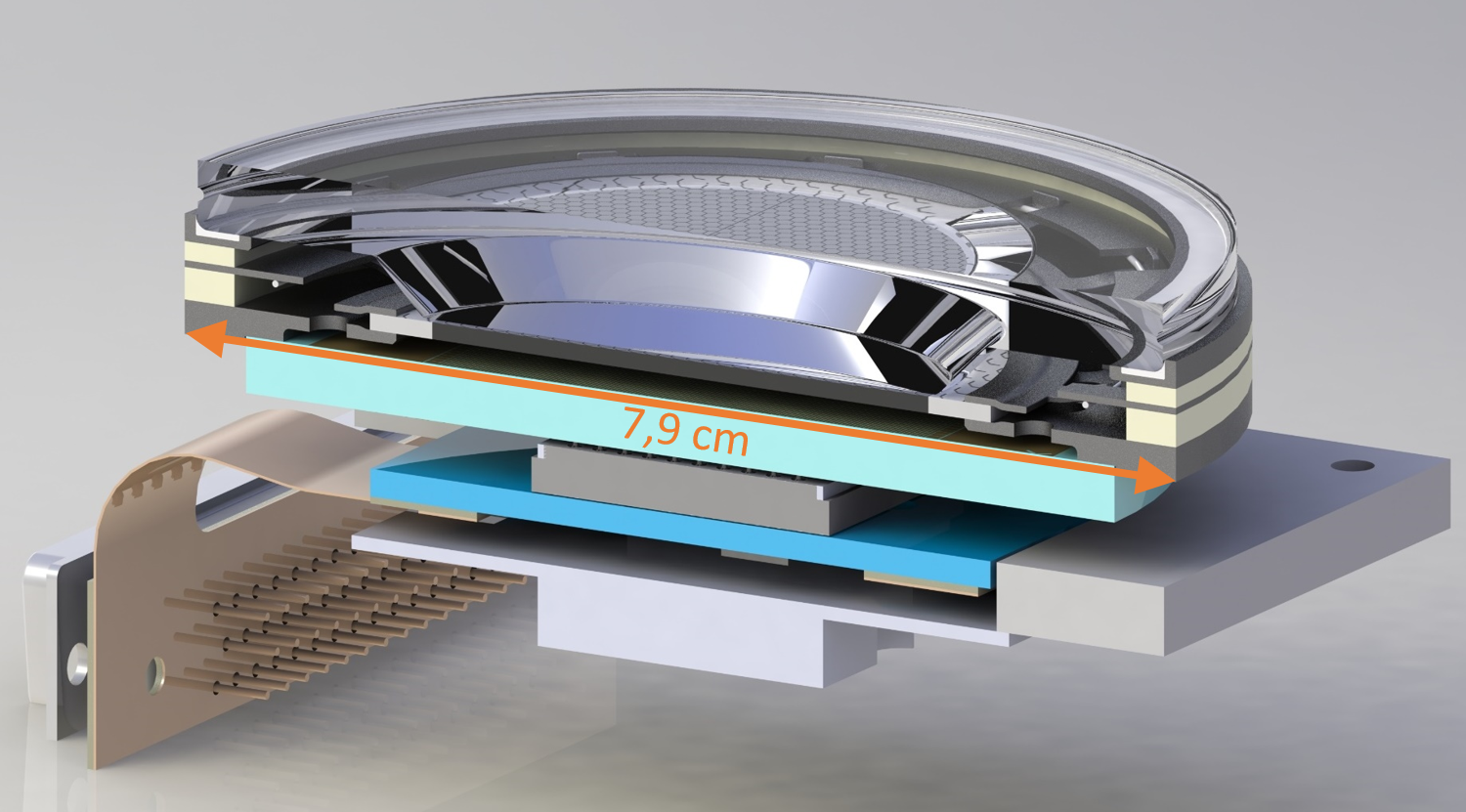}
\caption{Cross section of our detector} 
\label{fig:Detector_sectionalview}
\end{figure}

In this chapter we will look at the individual parts of our detector. A cross section of our detector is shown in Fig. \ref{fig:Detector_sectionalview}.

\subsection{Working Principle}\label{ss:working_principle}

\begin{figure}[t]
\includegraphics[width=\columnwidth]{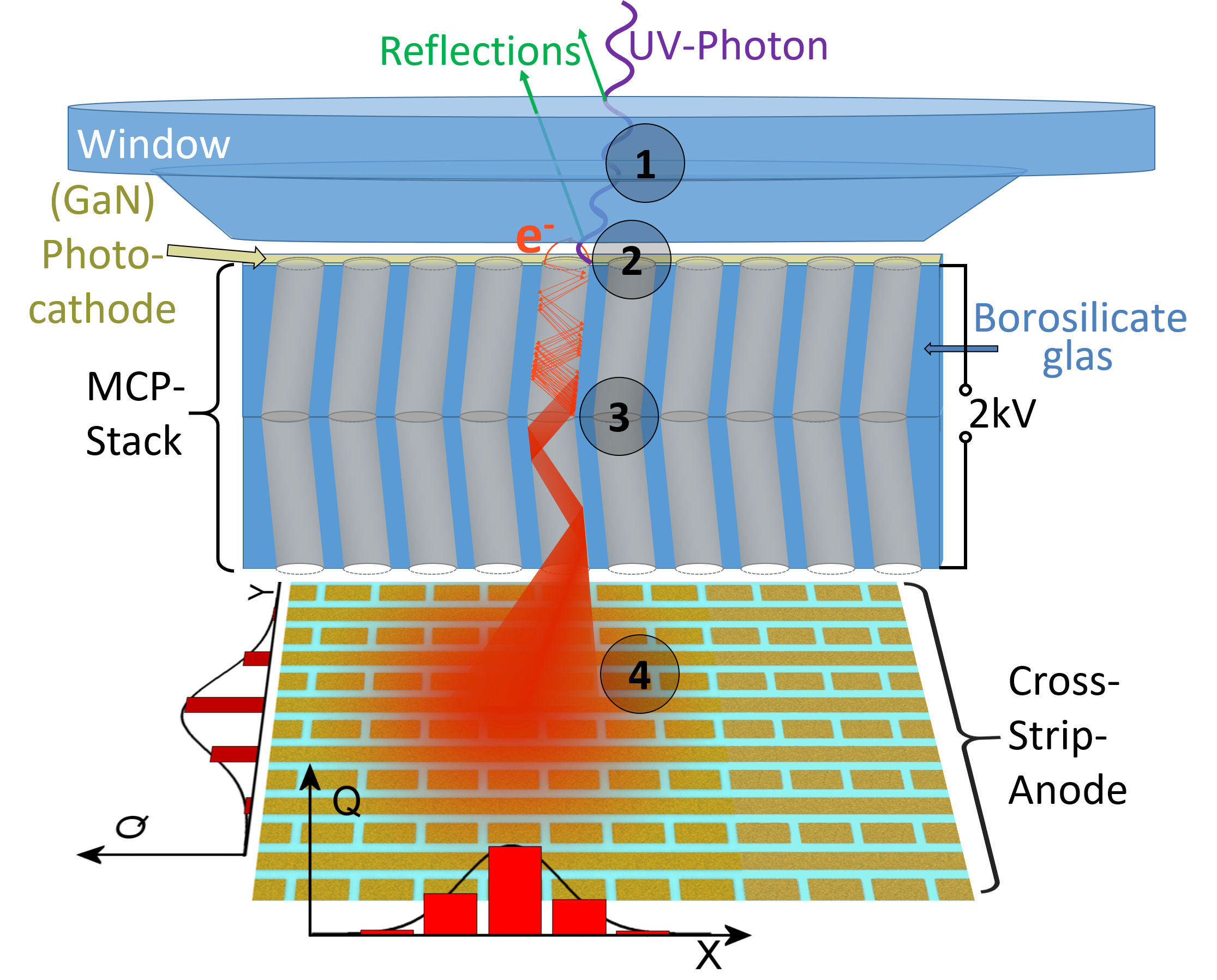}
\caption{Working principle of our detector (not to scale)} 
\label{fig:workingprinciple}
\end{figure}

The working principle of our detector will serve as an introduction to the more detailed view of the individual parts in the next chapters. The numbers in the following enumeration correspond to the numbers in Fig. \ref{fig:workingprinciple}.

\begin{enumerate}
	\item An incoming UV photon passes a \textit{window} (fused silica / MgF$_2$ / LiF) with low transmission from the far ultraviolet ($>$120~nm) to the near ultraviolet. 
	\item The \textit{photocathode} converts a photon into a photoelectron with a conversion rate depending on the photocathode material and the wavelength of the incoming photon. It can be as high as 70\% at 230~nm for GaN \citep{doi:10.1063/1.1883707}.
	\item In the \textit{microchannel plates} (MCPs) a high voltage of 2~kV (our HV supply might deliver up to 3~kV) accelerates the incident photoelectron, resulting in a charge cloud ($10^5$ to $10^6$ electrons) at the bottom side of the MCP stack. The position information of the incident photoelectron is preserved in the MCPs as the center of gravity of the electron cloud.
	\item The electron cloud is accelerated towards a \textit{cross strip anode} (64 strips in X direction and 64 strips in Y direction) and places a certain amount of charge on several strips of the anode.
\end{enumerate}

The charge signals on the individual strips of the anode consist of only a few thousand electrons and will be amplified to be readable by the front-end electronics.

This amplification is done by the \textit{BEETLE chip}, an ASIC with 128 charge amplifiers \citep{AGARI2004468}. It was developed by the Max Planck Institute for Nuclear Physics in Heidelberg for the LHCb experiment.

The output of the BEETLE chip is converted into digital signals by 4 ADCs and sent to our processing unit, a Virtex-5 FPGA. The implemented centroiding algorithm will calculate the center of gravity of the event and store it as one count in the corresponding pixel of the image.

The following chapters will present the crucial features of the individual parts listed above and the progress we made in these fields.

\subsection{Window and Photocathode}\label{ss:Design_window}

On the top of our detector, a window soldered to the detector body seals our detector to maintain an ultra-high vacuum (UHV) inside of the detector. This UHV is crucial for our cesium-activated photocathodes to be stable, as cesium reacts strongly with water and oxygen. The cesium in turn causes a negative electron affinity in the photocathode material. This is an important property of the photocathode to be able to emit electrons when absorbing photons.

To avoid the transmission through a window and to obtain spectral data in the far ultraviolet wavelength regime (FUV, $<$120~nm) it would be possible to use a door instead of a window. This so called "open design" would bring some disadvantages: it is heavier due to an opening mechanism that also could fail under space conditions and as the sealing of the valve won't allow a UHV inside of the detector, only suitable photocathode materials can be used.

In our current design and next generation design we chose to use a window. It has the disadvantage of having a certain cutoff frequency and an overall lower transmittance due to absorption and reflections \citep{doi:10.1063/1.1709744} but allows for the so called semitransparent mode. In this mode the photocathode is placed directly onto the window (further details: \cite{doi:10.1117/12.2054888}). As the window substrate is cheaper (in comparison to an MCP)  and can be reused after a proper cleaning process, quantum efficiency optimization measurements can be done with a large numberF of samples. We can use (001)-MgF$_2$ as window material for GaN. Its cutoff frequency (transmission at 7\%) at 110~nm \citep{0370-1328-72-6-308} makes it possible to include the Lyman-alpha line at 121.5~nm.

To achieve a flawless crystal growth of the photocathode material, a substrate material has to be selected which matches the lattice constant for one of the crystal planes and configurations. Furthermore, a difference in the thermal expansion coefficients can lead to tensions and cracks in the photocathode material.
To date we can produce Cs$_2$Te photocathodes with reasonable quantum efficiency as described in \cite{doi:10.1117/12.2054888}.
The next step is the growth and activation of GaN on MgF$_2$. The atoms of a MgF$_2$ crystal form a tetragonal crystal system with the P4$_2$/mnm space group \citep{weber2002handbook} which makes it possible to grow cubic GaN with a preferred (111) orientation on (001)-MgF$_2$ like shown in \cite{ganz2012wachstum}, chapter 5.2.1., at relatively low substrate temperatures of about $450-650^\circ$C. The production of the GaN layers on MgF$_2$ is accomplished by the research group of Daniel M. Schaadt. 

Further tests will show if we can grow GaN directly on atomic layer deposition MCPs (ALD MCPs). This would allow us to operate our detector in the more efficient opaque mode. Additionally, the opaque mode is more durable, as degassed ions from the MCPs are accelerated towards the window in semitransparent mode, leading to faster degradation of the photocathode material.

In principle the process was successfully tested by \cite{SIEGMUND2012168} and great progress has been made in this field as can be found in \cite{ERTLEY2017}.

\subsection{Detector Body}\label{ss:Design_body}

\begin{figure}
\includegraphics[width=\columnwidth]{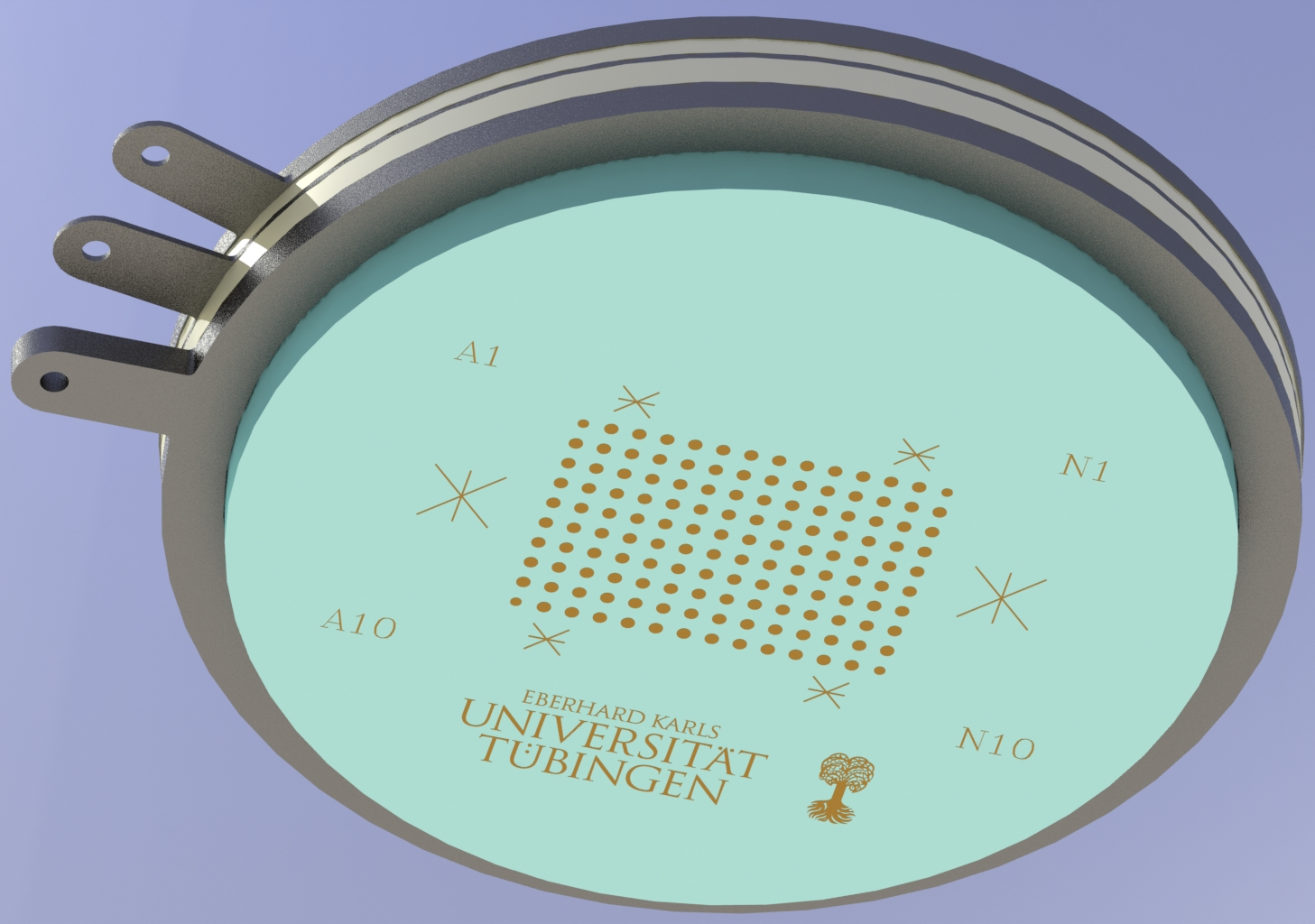}
\caption{Detector body and backside of the anode with its gold connectors and University logo} 
\label{fig:Detector_body}
\end{figure}

The detector body is composed of Kovar parts, insulating ceramic rings and the anode body at the bottom of the detector as shown in Fig. \ref{fig:Detector_body}. These parts are hard-soldered by H.R. Elsener (Empa), coated and tested for helium leak tightness. The tolerance in height should be below $\approx$100~$\mu$m for all five interfaces together, as the distance between the window and the first MCP is only 200-300~$\mu$m.

\begin{figure}[t]
\includegraphics[width=\columnwidth]{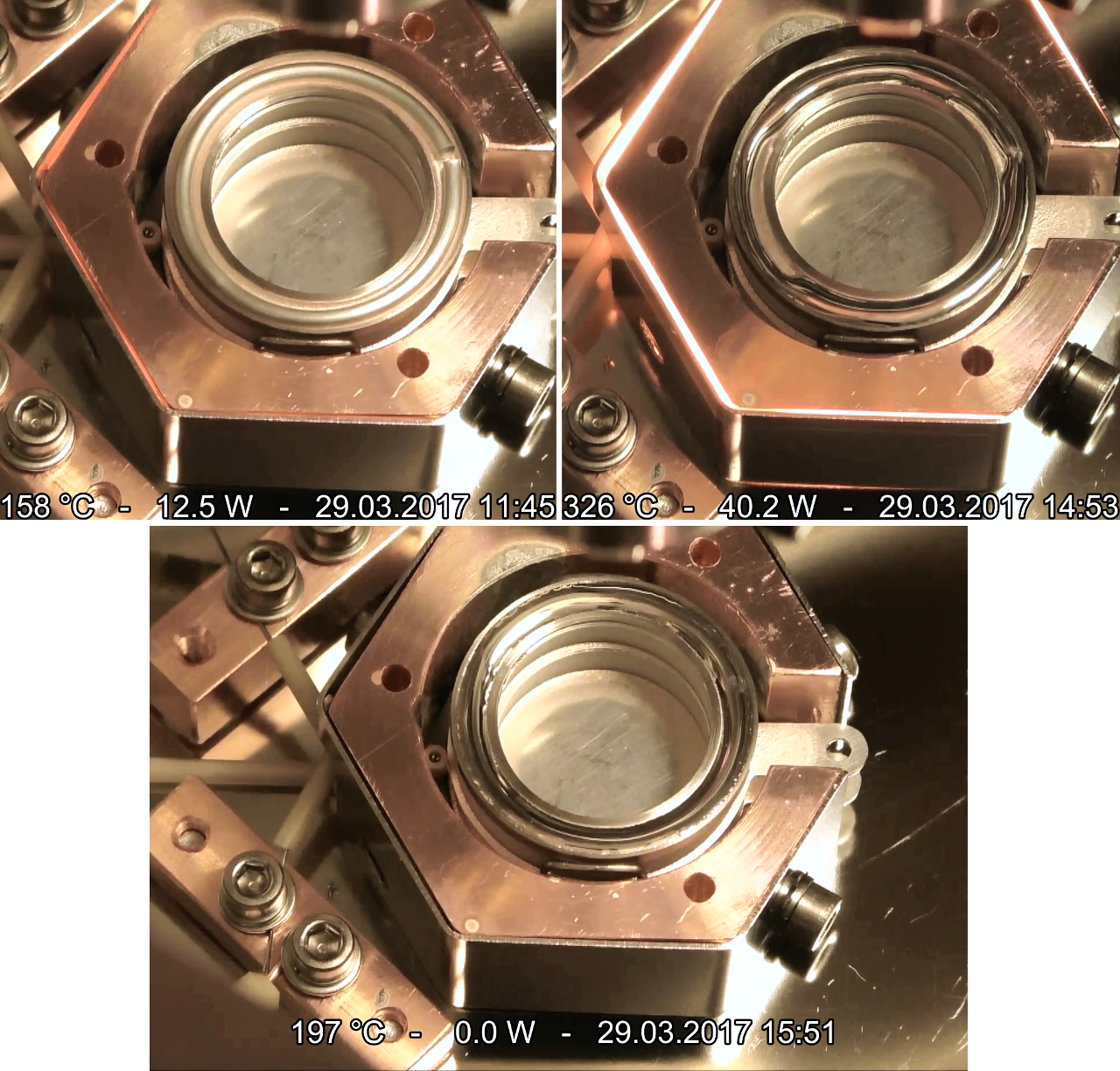}
\caption{Pictures from our sealing video documentation with current temperatures, heating power and time stamp as hardcoded subtitle. \textit{Top left}: Unmolten In-Bi ring (eutectic alloy). \textit{Top right}: Molten indium alloy at high temperature. Some parts of the sealing area are not covered properly. \textit{Bottom}: Indium alloy at lower temperatures covers the whole sealing area} 
\label{fig:Detector_sealing}
\end{figure}

The topmost Kovar part has some space where the indium alloy ring is placed right before the sealing process. In the sealing process, the detector body and window are heated up to about $240^\circ$C. The molten indium alloy ring should wet the whole sealing area. To achieve this goal, the surface tension between the indium alloy and all sealing parts should be relatively low. The surface tension between promising materials and the indium alloy was measured by H.R. Elsener (Empa) in a goniometer over the contact angle between the boundary of a molten indium droplet and a plate coated with the test material.
Our first tests with the new coating on the topmost Kovar part indicates a much better wetting of the sealing area as shown in Fig. \ref{fig:Detector_sealing}. As gold and a silver-copper alloy are some of the best materials found in this series of measurements, the sealing area of our window will be coated with a gold film.

To further improve the sealing process, we implemented a rotating stamp to twist the window while pressing it onto the detector. The purpose is to intermix the thin oxygen layer on the surface of the molten indium alloy into the indium alloy by this twisting. Further tests in the near future will show if our efforts in removing the oxygen layer from the indium alloy ring just before placing it onto the sealing surface and the wetting of the coated sealing surface by the molten indium on the coated sealing area are already sufficient to mix the oxide layer into the alloy.

\subsection{Microchannel Plates (MCPs)}\label{ss:Design_mcps}

\begin{figure}
\begin{center}
\includegraphics[width=0.75\columnwidth]{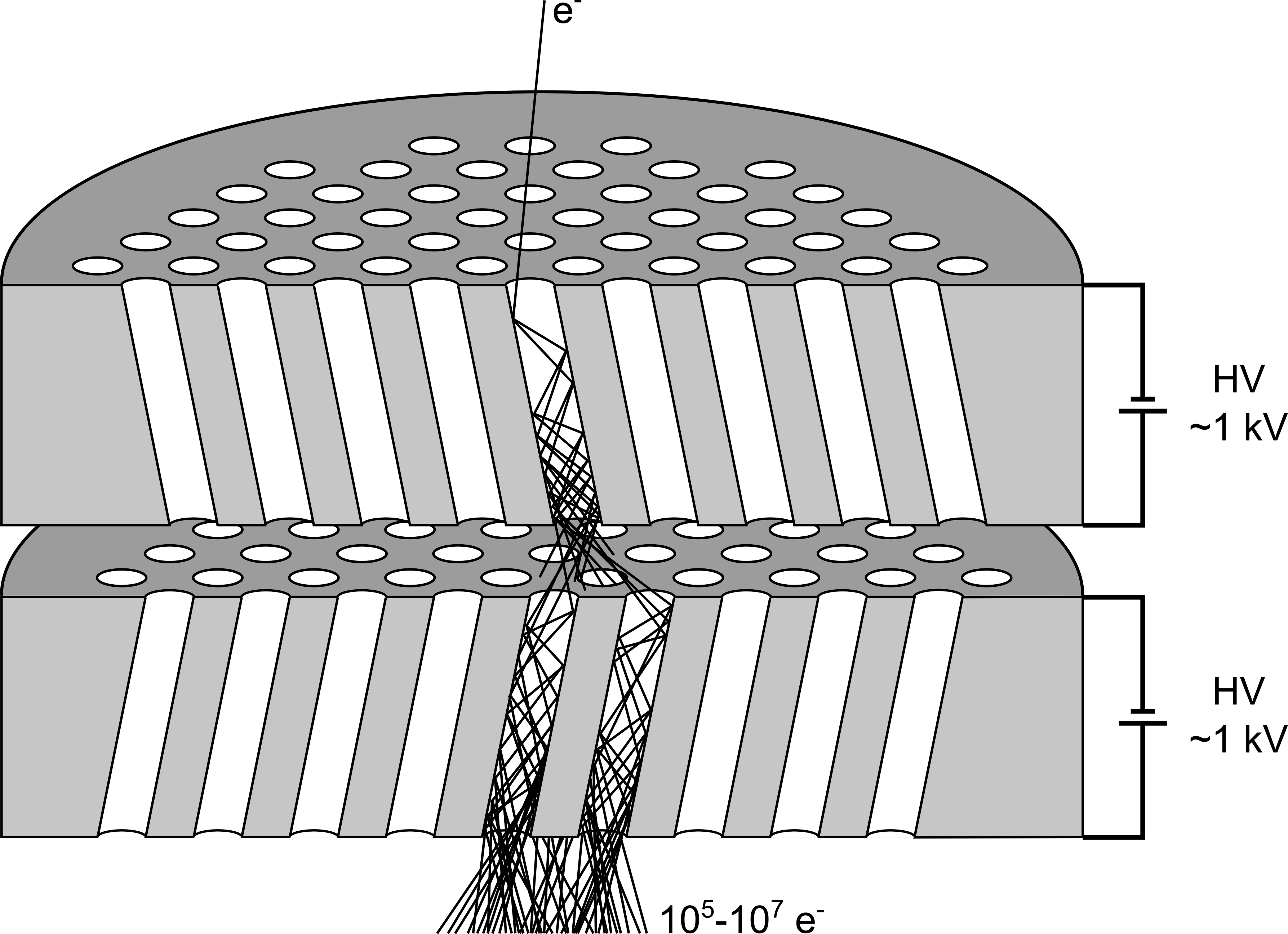}
\end{center}
\caption{Principle of the electron cloud formation in a stack of two MCPs as used in our detector} 
\label{fig:Detector_mcps}
\end{figure}


MCPs are used as a position preserving electron multiplier. The operational principle is illustrated in Fig. \ref{fig:Detector_mcps}.
Depending on the anode and readout electronics a gain factor of typically 10$^4$ to 10$^7$ electrons per photon is needed. At present, for testing the electronics we are using long-life MCPs (by PHOTONIS Technologies S.A.S., France/USA) with a relative low gain of 10$^5$. A smaller MCP gain results in a longer lifetime of the detector as the coating of the MCP channels degrades due to electrons hitting the surface of the channels. The characteristic variable for measuring the lifetime of the MCP is the total sum of charges per area and can be as high as $\tau_Q \approx$ 2--3 C/cm$^2$ for (non-ALD) long-life MCPs resulting in a lifetime of the MCPs of more than 10 years \citep{JINNO2011111}. 


For the next generation version of our detector we aim at atomic layer deposition MCPs (ALD MCPs) for three main reasons: 
\begin{enumerate}
	\item their lifetime is even longer than that of "classical" long-life MCPs as concluded in \cite{CONNEELY2013388},
	\item the degradation of the photocathode is reduced drastically as shown in \cite{CONNEELY2013388}, Fig. 6,
	\item as stated in \cite{ERTLEY2017} ALD MCPs withstand high substrate temperatures, needed to grow a GaN photocathode onto them or to implement a getter which needs high temperatures for activation.
\end{enumerate}
We did not choose to use ALD MCPs for our current design as the technology seemed immature at the time we started our current development. As resolution tests indicate that our optimal pixel size is about 20~$\mu$m (with an interpolation of 32 pixels/strip) the optimal pore size of the ALD MCPs is about 10-12~$\mu$m. These products are now commercially available (e.g. PLANACON\texttrademark \,by PHOTONIS Technologies) and first tests could already be done in 2018.

\subsection{Coplanar Cross Strip Anode}\label{ss:Design_anode}

\begin{figure}
\begin{center}
\includegraphics[width=0.75\columnwidth]{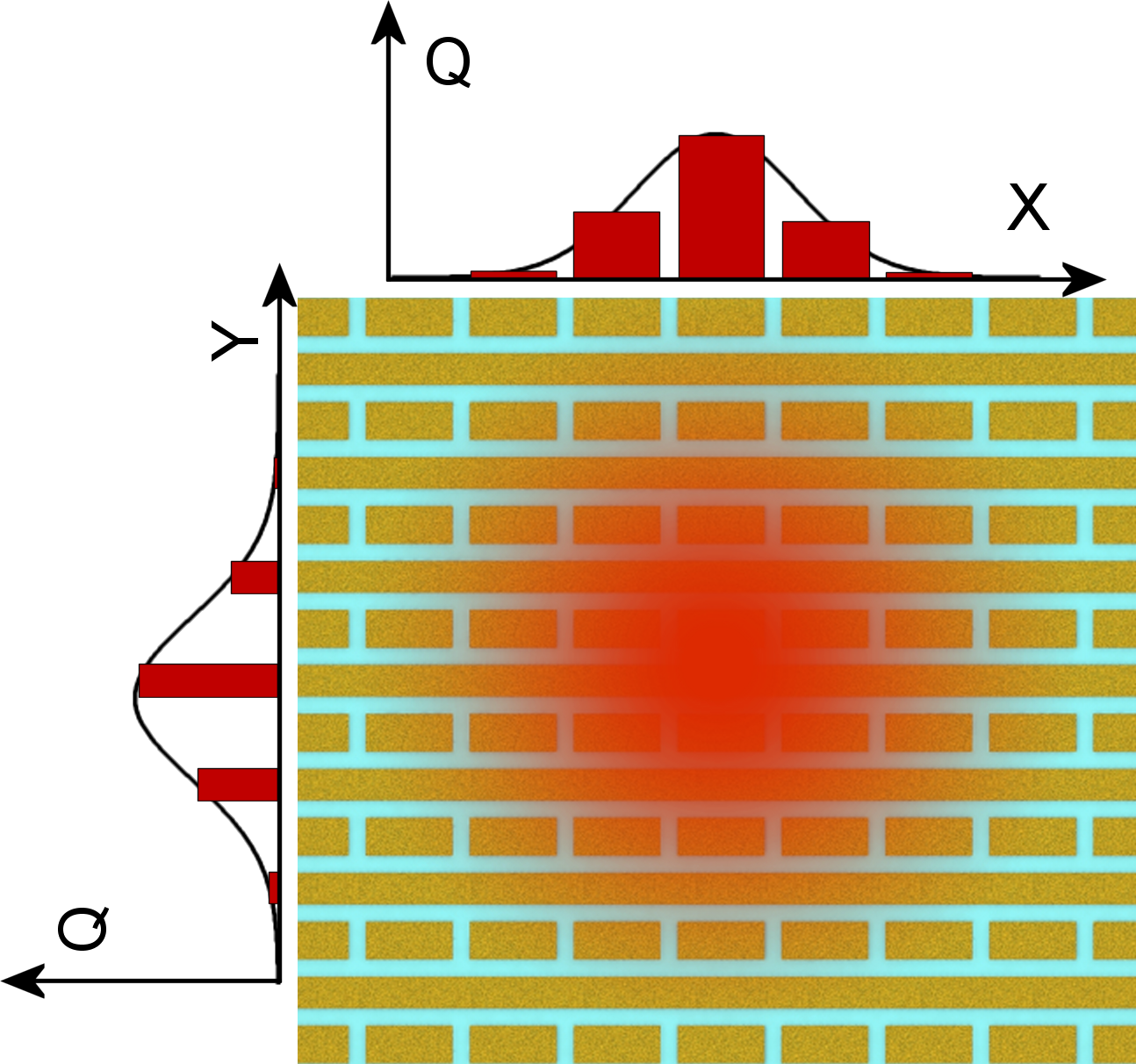}
\end{center}
\caption{Principle of the coplanar cross strip anode. Rectangles: vertically interconnected strips for the X direction. Horizontal electrodes: strips for the Y direction. Blue area: insulating anode body made from LTCC (Low Temperature Co-fired Ceramics). Red: simulated charge cloud depositing charges on several strips of the anode} 
\label{fig:Detector_anode_with_charge}
\end{figure}

\begin{figure}
\includegraphics[width=\columnwidth]{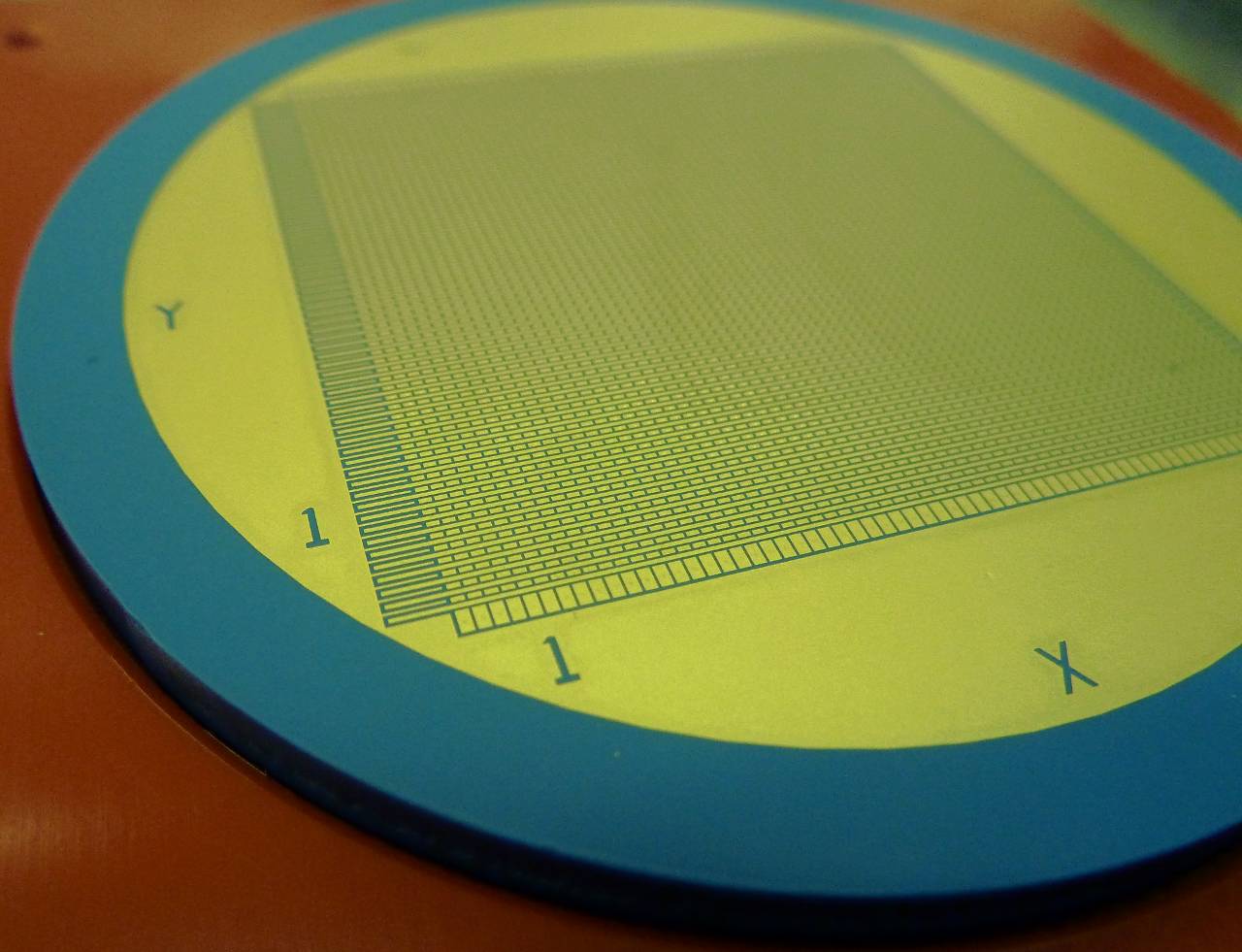}
\caption{Image of our anode. Golden parts: conducting electrodes (strips and grounding), blue parts: LTCC} 
\label{fig:Anode_image}
\end{figure}

The coplanar cross strip anode with 64 strips in each direction was designed by IAAT in cooperation with the manufacturer (VIA electronic GmbH, Hermsdorf, Germany). An image of the anode can be found in Fig. \ref{fig:Anode_image}. In contrast to the original design of cross strip anodes with two insulated layers of strips (see \cite{6488893}), our coplanar anodes are made from one surface electrode layer only. The originally lower strips are now composed of 63 individual electrode pads, which are connected by vias to a buried connection path within the ceramics body. This approach eliminates the need for an insulation layer and a second electrode layer atop of it. As shown in Fig. \ref{fig:Detector_anode_with_charge}, it has strips in the Y direction and rectangular fields for measurement of the X direction.

\subsection{BEETLE Hybrid Board}\label{ss:Design_beetle}

\begin{figure}
\includegraphics[width=\columnwidth]{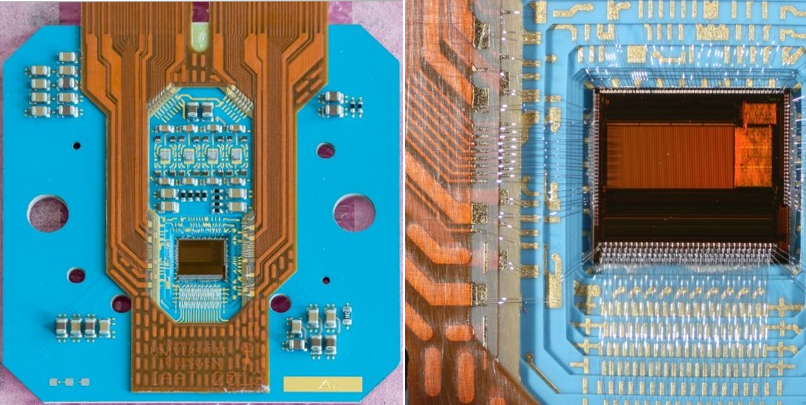}
\caption{Left: Image of the BEETLE Hybrid Board carrying the BEETLE chip. Right: magnified image of the BEETLE chip with its connecting bonds} 
\label{fig:Detector_anode_foto}
\end{figure}

The main purpose of the BEETLE Hybrid Board (see Fig. \ref{fig:Detector_anode_foto}) is to provide the highly demanding interconnections required to link each of the 128 anode strips to a corresponding input channel of the BEETLE chip. After the signals are preamplified and shaped by the BEETLE chip, the Hybrid Board also provides the outputs to send this data to our front-end electronics as well as the I$^2$C Interface to control the BEETLE chip parameters (further details in \cite{doi:10.1117/12.2054742}).

We recently added the possibility to control the BEETLE test pulses to our electronics. These test pulses are a feature implemented in the BEETLE chip for debugging and testing purposes. We found that the shape of these test pulses is dependent on the quality of the bonding between BEETLE chip and Hybrid Board and the connection to the anode strips, which makes it possible to characterize bonding errors and malfunctioning anode channels.

\subsection{Front-end Electronics}\label{ss:Design_readout_el}

\begin{figure}
\includegraphics[width=\columnwidth]{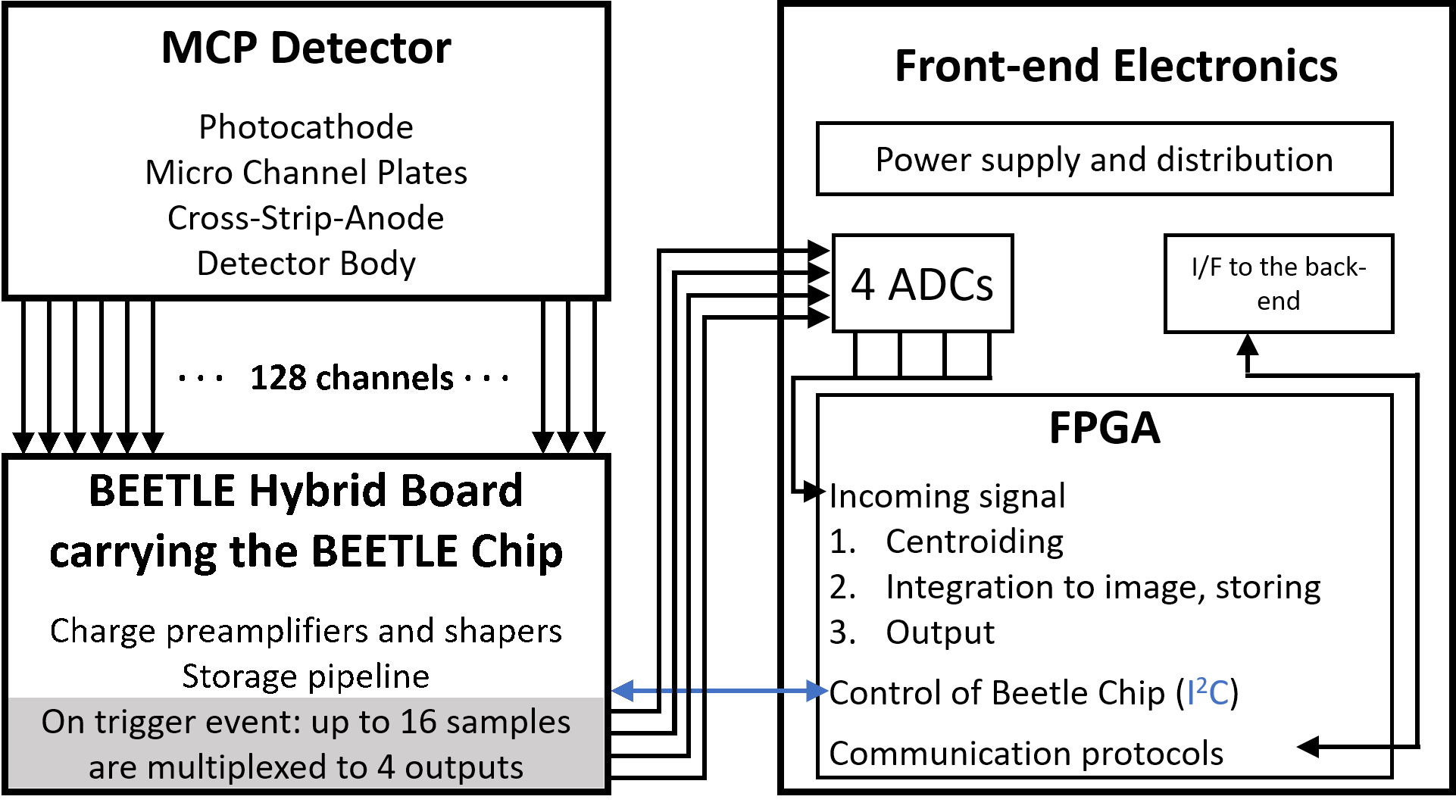}
\caption{Schematics of the detector and the readout electronics} 
\label{fig:Electronicsschematics}
\end{figure}

The readout electronics converts the preamplified pulses from the BEETLE chip into digital signals (via 4 ADCs) which are then further evaluated by our processing unit (see Fig. \ref{fig:Electronicsschematics} for the schematics of our readout electronics). The most demanding task is the centroiding of the charge signals. Each charge signal will be analyzed and stored as one count in the corresponding X-Y-position of our 2048x2048 pixel image, thus integrating an image from these position counts. In our final design we aim for a count rate of 300000/s, resulting in the need for highly sophisticated readout electronics. 

Right now, our data is analyzed offline on a workstation to test different centroiding algorithms (see chapter \ref{ss:data_evaluation_centroiding}). As these tests are about to be completed, we will encode the centroiding algorithm in hardware on a space-grade Virtex-5 FPGA.


As the Virtex-5 FPGA provides more than 800 user I/O pins, it is possible to implement the connection to the 4 ADCs, the backend computer, several SRAM modules (for the fast readout of lookup tables) and the I$^2$C bus.

\section{Data Evaluation}\label{ss:data_evaluation}


\subsection{Optimization of BEETLE operating parameters}\label{ss:beetle_parameters}

The pulse shape of the BEETLE charge amplifiers can be adjusted with several
parameters by programming the corresponding BEETLE registers (for details see \cite{Loechner:1000429} and \cite{lochner2006development} (Chapter 6.2)). This allows for optimization of pulse duration, amplification, pulse undershoot as well as signal to noise ratio and homogeneity of these values over all 128 BEETLE channels.

These values will depend also on the input capacitance, which may be different for individual BEETLE channels due to the more or less complex wiring from the anode electrodes through the anode ceramics substrate to the connector pads and then from the connector pads of the BEETLE Hybrid Board to the BEETLE inputs.

This optimization process is a multi-dimensional problem and therefore too complex to test all possible combinations. We have found already some promising settings, but this optimization process is currently still going on.

\subsection{Centroiding algorithms}\label{ss:data_evaluation_centroiding}

The goal of the centroiding process is to determine the center of gravity of the electron cloud. The readout electronics should be able to evaluate 300000 events per second. For that reason, we have to find a good compromise between computing time, complexity and accuracy. Since the shape of the electron cloud arriving at the anode is rather well described as Gaussian, a Gaussian fit might be the optimal solution regarding the accuracy. However, a fitting procedure comes at very high computation costs and complexity \citep{Delabie2014}. A promising algorithm is a 3 point Gaussian \citep{SUHLING1999393}. It takes the ADC value of the anode strip with the highest value $Q_0$ and the ADC values of the strips adjacent to it ($Q_1$ and $Q_2$). For these three given points the analytical solution for the center of the Gaussian is 

\begin{center}
$\overline{x}_{\text{Gau}} = \frac{  \text{ln} \left({Q{_2}}/{Q{_0}}  \right)}
                                  {2 \text{ln} \left({Q_1^2}/{Q_0 Q_2}\right)}$
\end{center}

\begin{figure}[t]
\includegraphics[width=\columnwidth]{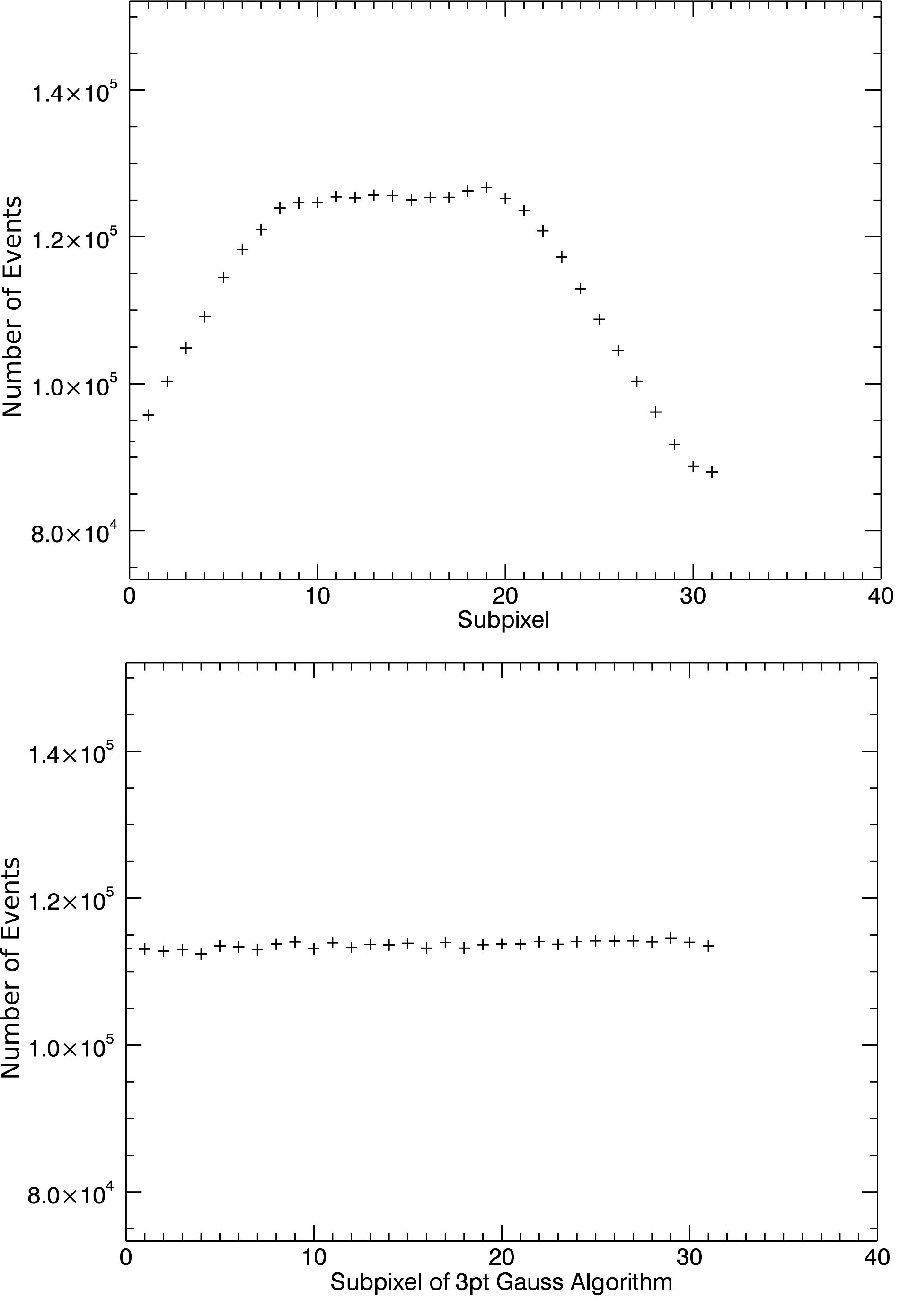}
\caption{The raw data of a flatfield image with 3.6 million events was analyzed with a 3pt Gauss centroiding algorithm. The number of events per interpolated pixel between two anodes should be constant in a flatfield image. \textit{Top:} uncorrected 3pt Gauss algorithm. \textit{Bottom:} 3pt Gauss algorithm with corrections} 
\label{fig:Electronics_event_rate}
\end{figure}

\begin{figure}
\begin{center}
\includegraphics[width=0.85\columnwidth]{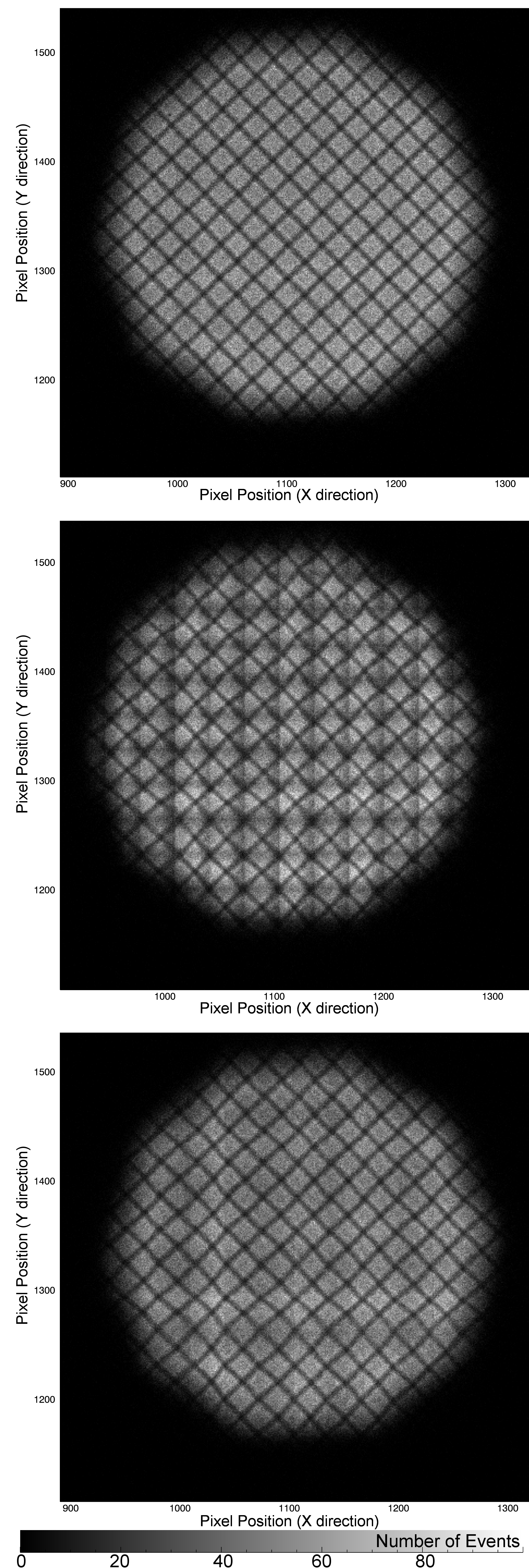}
\end{center}
\caption{Test images of a grid. Only about 1/5 in each axis of the image is illuminated. Three Gaussian centroiding algorithms evaluating the same raw data (3.6 million events) of a grid in front of the detector for comparison. \textit{Top:} Gaussian fit on each event (for comparison). \textit{Middle:} 3pt Gauss (without corrections). \textit{Bottom:} 3pt Gauss (with corrections)} 
\label{fig:Electronics_centroiding}
\end{figure}

With Q$_i$ normalized to Q$_0$=1, this algorithm can be implemented easily in an FPGA by using a two-dimensional lookup table. As shown in Fig. \ref{fig:Electronics_event_rate}, some corrections must be made to correct for the homogeneity of the distribution of the events over the interpolated distance between two strips. The resulting image is shown at the bottom of Fig. \ref{fig:Electronics_centroiding}.

Further improvements are necessary and could be achieved by the optimization of the BEETLE chip operating parameters and implementation of correction parameters for each anode strip.

\section{Summary and Outlook}\label{s:summary}

We presented our space-flight MCP detector with its low power consumption (10-15~W), compact size and light weight ($<$3\,kg for detector, electronics and HV). As we can change the photocathode material and switch between open and closed detector designs, we can fit our detector to specific missions.

We are developing two versions of our detector. The current design uses Cs$_2$Te photocathodes in semitransparent mode and long-life MCPs. As the real-scale detector bodies with a diameter of 7.9~cm are about to be finished, we are able to test our first prototype within 2018. The last steps to finish the current design are to scale all processes to the real-scale detector bodies, test our sealing process, implement our centroiding algorithm and migrate our FPGA code onto a Virtex-5 FPGA.

The next generation design for the European Stratospheric Balloon Observatory Design Study will use GaN photocathodes (or similar photocathode materials with respect to QE) and ALD MCPs. We will use windows with high transmission in the FUV (wavelength cutoff $\leq$ 120~nm) to finally obtain as much spectral data as possible from astronomical objects visible in the UV.

\section{Acknowledgments}\label{s:acknowledgments}
This work is funded by the Deutsches Zentrum f\"ur Luft- und Raumfahrt (DLR) under grant 50\,QT\,1501.

%
%

%
%

%

%
%

%

%


\emergencystretch=2em

%
\bibliographystyle{spr-mp-nameyear-cnd}  
\bibliography{biblio-u1}                

\end{document}